\documentclass[aps, twocolumn, showpacs]{revtex4-1}
\usepackage{amsmath}
\begin{document}
\title{Physics and mathematical reality: comments on Aharonov et al, arXiv: 1902.08798}
\author{S. C. Tiwari \\
Department of Physics, Institute of Science, Banaras Hindu University, Varanasi 221005, and \\ Institute of Natural Philosophy \\
Varanasi India\\}
\begin{abstract}
Do we have two kinds of reality: physical and mathematical? What is the role of mathematics in physics? These fundamental questions
have intrigued original and brilliant minds since ancient times. A recent article (Aharonov, Cohen and Oaknin, arXiv: 1902.08798)
offers a viewpoint on the use of abstract mathematics in quantum theory. They argue that, in some cases, it may result into a paradox 
focusing on an example in quantum mechanics regarding Aharonov-Bergmann-Lebowitz (ABL) rule. In this comment 
we discuss three issues: theory reduction and approximation, origin of a paradox, and the reality of time. It is pointed out that paradoxes
appear even in mathematics; we explain them as the results of the acceptance of pure logical artifacts. In physics too, the impact of the so called
quantum and relativity revolutions, have led to the emergence of counter-intuitive and mysterious notion of the physical reality. The most
fundamental change has been on rejecting the reality of time. A radically new perspective 
that physics is the language of mathematics is suggested.
The mathematical reality of the natural numbers manifests in the form of physical time.
 
\end{abstract}
\maketitle

\section{\bf Introduction}
The significant role of mathematics in physics has been recognized for past hundreds of years, more sharply after Newton's 
Principia: in his own words his work represents ``the mathematical principles of philosophy''.
Poincare presents profound thoughts on the mathematical truth and the age of mathematical physics. Relatively recently prominent theoretical 
physicists articulated and debated the physics-mathematics
relationship with differing perceptions \cite{1}. It is usual to state that mathematics is the natural language of physics, for example, Schwinger
notes this in his Nobel lecture. Pure mathematics would then appear alien to the physical world: mathematical truths exist somewhere even outside the
mind of a mathematician to be discovered by  him/her. Hardy seems to believe in this kind of pure mathematics \cite{2}. He admits
that philosophers and mathematicians have differing concepts on mathematical reality. His own position stated 'dogmatically'
is, 'I believe that mathematical reality lies outside us, that our function is to discover or observe it; the theorems as creations are simply 
notes of our observations'. To some physicists, only utility for the
description of physical phenomena would determine the value of mathematics. I think the perspective adopted by Aharonov et al \cite{3} on mathematics
seems to support this narrow viewpoint. The concluding part of the revised version of the article \cite{3} asserts:
``This analysis may support a general viewpoint according to which when addressing a physical problem one should first understand the governing 
physical principles, as well as the involved approximations and assumptions, before heading to the mathematical formalism.
Otherwise, a misguided use of the formalism could lead to apparent paradoxes. In such cases one should return to the fundamental physical principles 
in order to trace the origin of the paradox and then solve it by employing a mathematical model which better accords with
these principles. All this, of course, does not diminish the major importance of mathematical
language in describing physical problems, but only calls for a careful use thereof which is aligned with the physics.''

This conclusion needs deeper analysis. Authors \cite{3} raise the question related to the use of abstract mathematics in quantum theory. They 
point out that in some cases this could lead to paradoxes: in a specific example of the calculation of conditional probablities
using the Aharonov-Bergmann-Lebowitz
(ABL) rule. Three important issues are involved here: 1) theory reduction, 2) the origin of a paradox, and 3) the nature of time. The authors \cite{3}
remark that, 'It is important not to forget the physical understanding of the assumptions and approximations involved in the mathematical formalism'.
One would agree with it, however, the meaning of physical understanding itself could be controversial, specially so in the case of quantum theory.

The meaning of reality, both physical and mathematical, and the role of mind in the perception of the reality are fundamental questions \cite{4}.
In contrast to Hardy's belief, Leibnitz had mystic interpretation, in the words of Laplace: 'He (Leibnitz) imagined that unity represented God,
and zero the void; that the Supreme being drew all beings from the void, just as unity and zero express all numbers in his system of numeration'.
Kronecker is quoted to have said,'God created integers, the rest is the work of men'. I refer to a charming book written for 
'the cultured non-mathematician' according to the author Dantzig \cite{5}. We have discussed Poincare's ideas in \cite{4}. 
He says that mathematics overlaps both physics and philosophy, however
it does not take much from the outside world. Mathematical mind can create abstract things if freed from the 'tyranny of the external world'. Nevertheless
mathematicians need to solve the problems of physics arising from the external world. Experiment is the only source of the physical truth. Intuition
is the instrument of mathematicial creation, however logic gives rigor to it. 

Now, Dirac also offers insight into the place of mathematics in physics that is
nicely explained by Yang in \cite{1}, pages 501-502 and Figure XII.1. According to Dirac there are two routes to the mathematical concepts 
useful to fundamental physics. First one is the organization of experimental information into a physical law inventing mathematical concepts. 
The second route is to begin with beautiful mathematics and relate it to experimental reality. Newton's rules of reasoning for experimental
natural philosophy essentially represent the first route. Dirac (and Yang) prefer the second route in the 
light of enormous complexities of modern experiments. 

In this note we offer a constructive criticism on the motivation and the message of the authors in \cite{3} and assert that the concept of time
is the most important fundamental issue in the foundations of physics and mathematics. In the next three sections we discuss aspects of theory 
reduction, paradoxes and time respectively. In the last section the new viewpoint on time is articulated to stimulate further exploration of this idea.

\section{Theory Reduction}

Validity domain of a physics theory, its generalization, and equivalence of two theories have intriguing and interesting aspects. Rohrlich \cite{6}
elucidates theory reduction limited to formalisms. Following him, let us consider a theory T approximated by theory S, then reduction is a 
procedure to obtain S from T. Usually it is just the formalism of T that under suitable limits reduces to the formalism of S. Rohrlich
remarks that the semantic component of T may not necessarily be related to that of S. Here semantic element is taken to be the 
interpretation or physical model in \cite{6}. We have argued \cite{4} that a complete theory reduction involves three ingredients: mathematical
formalism, physical interpretation, and foundations. To take an example, consider T to be the theory of general relativity and  reduction to Newtonian 
gravitational theory. In
this case, only partial formal reduction is obtained as the physical interpretation also gets changed. The foundations of two theories, on
the other hand, have nothing in common. As for the inequivalence of interpretation of two theories, Lorentz transformations are formally same in
special relativity and Lorentz theory, but Lorentz treated time as mathematical having no physical significance in the transformations. In
fact, special relativistic Lorentz group to Galilean group reduction is just formal; the time coordinate has entirely distinct meaning in the two cases.
Matrix mechanics and wave mechanics in quantum theory were proved formally equivalent by Schroedinger himself. However matrix mechanics
is founded on the concept of relations between observables discarding the space-time representation opposed to the wave mechanics \cite{7}.

The Nobel Lectures by Tomonaga, Feynman and Schwinger indicate that their approaches had differing conceptual perspectives on the development of
quantum field theory though all the three were inspired by Dirac's ideas. Schwinger states that, 'Indeed,
relativistic quantum mechanics - the union of the complementarity principle of Bohr with the relativity principle of Einstein - is
quantum field theory'. He notes that his approach is differential whereas that of Feynman is integral; both are based on 
Dirac's transformation theory and classical
action. Note that the role of time discussed by Schwinger and Feynman has curious differences, and that of Tomonaga is markedly different. Dirac's many-time
theory, i. e. N space coordinates and N time coordinates for N particles, inspired Tomonaga; in his own words ``This paper of 
Dirac's attracted my interest because of the novelty of its philospohy and the beauty of its form''.
It is true that after Dyson's 1949 papers on Schwinger-Tomonaga theories, and later the proven utility of path integral approach
such foundational issues on QFT do not seem to have practical use. However, I think the significance of time in QFT has to be examined afresh. Specifically,
space-time localized measurements, and causally independent measurements in space-like regions in Schwinger's theory, and many-time concept in
Tomonaga's theory have to be carefully analyzed.

\section{Paradoxes}

Aharonov et al \cite{3} underline the necessity to trace the origin of a paradox, and for this purpose suggest the significance of fundamental
physical principles. Unfortunately, in quantum theory the physical interpretation of fundamental principles continues to remain unsettled: theory
of measurement and wavepacket reduction are two of the most important such principles. Before attributing the origin of a paradox 
to the use of abstract mathematics or information theory, a brief digression to paradoxes in mathematics
would be illuminating.

The Cantorian antinomies in the 19th Century created sharp divisions among mathematicians leading to the emergence of prominent opposed schools of thought:
intuitionists and formalists. For finite sets, the differences were only in their methods, but for the infinite sets there were 
fundamental differences. A set is enumerable if it can be put into one to one correspondence with the set of natural numbers. What
is the total number of the elements of such infinte sets? Cantor introduced the concept of transfinite numbers. These denote the 
totality of the elements of such sets, and for natural numbers it is called aleph-null, the power or potency of the set of natural numbers. 
In this theory aleph-one is greater than aleph-null. It could be proved that rational and algebraic numbers are enumerable, and also have 
the power aleph-null. It is not known if the power of the set of real numbers is aleph-one: the problem of the continuum. In Cantor's
theory there is infinity of transfinite numbers. One may ask whether the infinity is real. Intuitionists reject the actual reality of the 
mathematical infinity, and argue that transfinite numbers are non-constructible. Poincare suggests
the origin of antinomies to the assumption that the infinite is actual real. Formalists, on the other hand,
believed that the contradictions originated due to the 
lack of a precise language: symbolic language, and formalization of mathematics based on axioms, rules of inference and formal deductions became
their goal. This programme suffered a setback after the Goedel's theorem or results. Though Goedel considered only formalized arithmatic system 
the conclusions of the Goedel results were of general validity.

Subsequently Turing introduced the idea of computable numbers and a mechanical model of algorithmic procedure in what has come to be known as
Turing machine. The importance of information theory in computer science, and the argument that information is physical have led to the assertion
that, 'mathematics and computer science are a part of physics' \cite{8}. Algorithmic complexity theory has been used to propose that only
``humanly meaningful number set'' is acceptable \cite{9}. Theory of computable numbers is first suggested to be of physical nature. and 
in the next step Church-Turing hypothesis is promoted to a principle as Deutsch puts it: Every finitely realizable physical system can be 
perfectly simulated by a universal model computing machine operating by finite means. It has been argued that classical physics is false, and
that quantum theory is compatible with the Church-Turing principle. The emergence of quantum information science has led to a school of
thought that mathematical reality is a part of physics and that physics is a branch of information science. I believe this kind of the 
nature of reality is superfluous that may appeal the world of artificial intelligence, not a true mathematician or a true physicist concerned
with the foundations.

Let us recall that the crisis in the foundations of mathematics occasioned by Goedel's theorem led to the realization
that mathematical intuition is of fundamental importance in resolving the paradoxes, and that contrary to the belief of formalists the 
language was not responsible for the origin of the antinomies. Learning from this I think the resolution of two conceptual crises in physics emanating
from relativity and quantum theory should have followed a radically different path founded on physical intuition not counter-intuitiveness. Further, if
mathematics is the language of physics it possibly could not result into paradoxes. Therefore I suggest that the origin of paradox
attributed to abstract mathematics in \cite{3} is not convincing: it could be due to the interpretational problem of quantum theory. In the next
section we explore this idea.

\section{Time Asymmetry}

The problem discussed in \cite{3} is that of finding the probability for a physical observable in a projective measurement at
an intermediate time t between initial time $t_i$ and final time $t_f$ for a unitary time-evolution of a quantum system governed
by a time-independent Hamiltonian. Equation (3) in this paper \cite{3} defines the conditional probability for the assumed initial
and final quantum states. Relation (3) has time-symmetry even for the collapse of the wavefunction. For a particular case of
physical observable an apparent paradox is noted by them. The authors argue that the ABL rule defined by their relation (3) is
restricted to the projective measurement. One has to incorporate the uncertainty in the measuring pointer and use positive-operator
valued measure for the generalized ABL rule to resolve the paradox.

In this case-study the interpretation of two principles is involved, namely, the meaning of quantum measurement and time symmetry. Note that
one of the views asserts that the basic interpretation problem in quantum mechanics is the question as to how the superposition of
quantum states and the collapse of the wavefunction upon measurement are reconciled. von Neumann showed \cite{10} two kinds of quantum
measurement: infinite regression and discontinuous reduction. A recent paper \cite{11} dwells on the issue of weak values, projective
measurement and weak measurement. An interesting idea in Section IV of \cite{11} is that of the transition path time distribution. We may
ask if these works \cite{3, 11} could be placed in the context of von Neumann's two kinds of measurement.

Two alternative scenarios having bearing on this issue are worth mentioning. An abstract mathematical idea in the 
form of Weyl-Kaehler space \cite{12} offers a new perspective on the physics of quantum measurement.
Briefly stated the proposition builds on the consistent histories approach for a closed system. One postulates a projector $Y$ on the 
history space $\overline{H}$
\begin{equation}
 \overline{H} = H \otimes H \otimes ........\otimes H
\end{equation}
\begin{equation}
  Y = E_1 \otimes E_2 \otimes ........\otimes E_n
\end{equation}
Here a sequence of events at successive times $t_1 < t_2 .....<t_n$ is represented by projectors $E_1, E_2, ....E_n$ on the Hilbert
space, and the history space is a tensor product of n copies of $H$. A consistent Boolean algebra of history projectors is termed a framework.
A priori time-order is not needed in this approach.

The new element in our approach is the geometry of quantum state space generalized to Weyl-Kaehler space in which the splitting of wavefunction
upon measurement has natural explanation. The tensor product space (1) is replaced by a multiply-connected space and topological index defines
a time parameter. A state vector $|\psi>$ on the Hilbert space $H$ jumps to other connected Hilbert space and becomes $|\phi>$. Postulating
\begin{equation}
<\phi | \phi> = e^{n \Lambda} <\psi | \psi>
\end{equation}
each successive measurement is characterized by index $n$, and exponential factor is a weight factor. Integer $n$ defines a time-ordering
of the measuring events. Here
\begin{equation}
\oint A_i dz^i = n \Lambda 
\end{equation}
\begin{equation}
| \psi > = | z_0, z_1 ..... z_N> 
\end{equation}
Since $A$ is closed, the quantum measurement is interaction-free process similar to the Aharonov-Bohm effect. 
Collapse of wavefunction is interpreted as a transition of the state vector from one Hilbert space to another.

The second scenario has fundamental significance for both quantum mechanics and special theory of relativity: a paradox related with
the lifetime of an unstable particle traveling at relativistic speed \cite{13}. First we note the empirical fact that the decay rate
of an unstable system is described quite well by exponential decay law. It is also experimentally observed that lifetime of a particle moving at
a relativistic speed is longer than that of at rest.. Decay is a time-irreversible process. How does time-symmetric quantum theory become
applicable in this case?  Why should relativistic time dilatation kinematic effect account for the observed lifetime of a fast moving particle?

Excerpts from \cite{13} presented in the Appendix bring out the conceptual problems and the origin of the paradox. In addition to this
a brief comment on Weinberg's treatment \cite{14} of the problem is made here. Weinberg remarks that the theory is developed based on the way
the experiments aew done. In a collision process particles are assumed to have localized wavepackets far from each other, and the time-history
of the superposition of these states is followed. The interaction lasts for a short time interval $T$, and a time-box is assumed. The energy
consevation assumes the form
\begin{equation}
\delta_T (E_\alpha -E_\beta) = \frac{1}{2 \pi} \int_{-T/2}^{T/2} e^{i(E_\alpha -E_\beta) t} ~dt 
\end{equation}
After few steps a master formula for S-matrix is obtained; see Equations 3.4.11 and 3.4.12 in Section 3.4 of \cite{14}. An interesting special case is 
for single-particle state $\alpha$ decaying into a multi-particle state $\beta$. The Lorentz transformation is discussed arriving at the 
conclusion stated on page 138 that, 'the decay rate has the same Lorentz transformation property as $\frac{1}{E_\alpha}$. This is, of course,
just the usual special relativistic time dilation - the faster the particle, the slower it decays'.

Weinberg seems to resolve the paradox in which both quantum mechanics and relativity have been used. However, a closer look is necessary critically
examining the S-matrix theory outlined in Chapter 3 of \cite{14}. The salient points could be stated as follows. 1) Interaction occurs at small
microscopic length scale while the observations are made at macroscopic distances. Thus the cross-sections measure the transitions between
non-interacting distant 'in' and 'out' states $\psi^{\pm \infty}$. 2) The states $\psi^{\pm \infty}$ are defined at times $t=\pm \infty$.
For manifest Lorentz invariance the state vectors are assumed to be independent of time in the Heisenberg picture. As a consequence the state 
vectors 'in' and 'out'  are not the limits at $t \rightarrow \pm \infty$ of a time-dependent state vector. They do not belong to two different
Hilbert spaces. And, 3) the S-matrix contains Dirac delta-function $\delta^4 (p_\alpha -p_\beta)$ to account for the energy-momentum
conservation. In the cross-sections or transition rates one has squared S-matrix, and the square of distribution or Dirac delta-function 
is not quite well-defined. For this purpose approximations are introduced in Section 3.4 of \cite{14} to make physical sense of the squared S-matrix.

Has the ambiguity in quantum mehanics been compensated by the ambiguity in relativity on the physical interpretation of the concept of time?
It seems the state vectors defined in different inertial frames, Dirac delta-function and time-box, and the Lorentz invariance of the S-matrix 
somehow conspire to result into the Weinberg's conclusion stated earlier regarding the time dilatation of lifetime. What is
time-operator in quantum mechanics? How does one define inertial frame quantum mechanically? Does lifetime behave as a clock in an inertial frame?
What is a quantum clock? To my knowledge a consistent satisfactory answer to all these questions has never been given.

The present trend in one of the influential school of thought discards the reality of intuitive time, and seeks interpretation of
space-time as emergent based on abstract Hilbert space of quantum states or on twistor space. A speculative conjecture ER=EPR has become fashionable
in mainstream physics literature. This trend would lead to paradoxes and conceptual crisis in physics; it may create artificial physical reality. We
have pointed out that \cite{15} the main reason for the apparent dead-end in superstrings is the side-lining of truth and beauty in the
conception of physical reality.

\section{\bf Conclusion}

Intuitive reality of time is linked with the mind in the last paragraph of \cite{13} and the soul of the monograph \cite{4} Chapter 4 
proposes: ``Time creates time, time is action itself as well as a measure of change'' and the natural numbers find expression in the flow of time.
We put them in the form of a new hypothesis: Physics is the natural language of mathematics. The mental creations in mathematics find nearest
expression in the tangible world of theoretical physics. For example, the incessant flow of time is an approximation to the natural 
numbers and the law of induction. Physical circular objects approximate the geometric circle.

A logical extension of our hypothesis is to find a method that links topology and geometry with physics. Let me recall an interesting
mystic observation from \cite{5}. The Euler identity (de Moivre's formula) $e^{i \pi} + 1 =0$ may be interpreted as unification of 
arithmatic (0 and 1), algebra (symbol i), geometry ($\pi$) and analysis (e).

The mystery in the Schroedinger equation is due to the presence of $i$: it spoils the analogy to the diffusion equation \cite{7, 10}.
However Nelson's stochastic approach \cite{16} develops the theory with the probability as basic element. In an important step in this 
approach de la Pena et al \cite{17} show that quantum mechanics is a partially averaged approximate asymptotic theory following from 
Fokker-Planck-type equation in phase space. Could we re-visit Schroedinger equation replacing $i$ by something to get real functions?
The discussions in \cite{7, 10, 16, 18} show the importance of this objective. An interesting formal proposition is to make following replacement
$ i \rightarrow \begin{bmatrix} 0 & 1 \\ -1 & 0 \end{bmatrix}$. The Schroedinger equation transforms to a set of coupled equations
for the doublet $\begin{bmatrix} \phi \\ \psi \end{bmatrix} $
and could be combined to give fourth-order equation originally derived by Schroedinger, see \cite{18}. Do probability/statistical
methods relate physics with mathematics? We are exploring this question to seek relationship of topological origin of spin of electron \cite{19}
to that from the stochasticity \cite{20}.


\begin{thebibliography}{99}
\bibitem{1} Some Strangeness in the  Proportion, Edited by Harry Woolf (Addison-Wesley, 1980)
\bibitem{2} G. H. Hardy, A Mathematician's Apology (C. U. P. 1967)
\bibitem{3} Y. Aharonov, E. Cohen, and D. H. Oaknin, Why physical understanding should precede the mathematical formalism - conditional 
quantum probabilities as a case-study, arXiv: 1902.08798 v2
\bibitem{4} S. C. Tiwari, Time-Transcendence-Truth, IONP Studies in Natural Philosophy Volume 1 (lulu.com, 2006)
\bibitem{5} T. Dantzig, Number the Language of Science (George Allen and Unwin Ltd, 1942)
\bibitem{6} F. Rohrlich, There is good physics in theory reduction, Found. Phys. 20, 1399 (1990)
\bibitem{7} S. C. Tiwari, Derivation of the Hamiltonian form of the Klein-Gordon equation from Schrödinger-Furth quantum diffusion theory: Comments,
Physics Letters A 133, 279 (1988)
\bibitem{8} R. Landauer, The physical nature of information, Phys. Lett A 217, 188 (1996)
\bibitem{9} J. Ford, How random is a coin toss?, Phys. Today, 36(4), 40 (1983)
\bibitem{10}  M. Jammer, The Philosophy of Quantum Mechanics ( John-Wiley, 1974)
\bibitem{11} E. Cohen and E. Pollak, Determination of weak values of quantum operators using only strong measurements,
 Phys. Rev. A 98, 042112 (2018) 
\bibitem{12} S. C. Tiwari, Geometry of quantum theory: Weyl-Kaehler space, arXiv: quant-ph/0109048, Published in Geometry, Analysis
and applications, edited by R. S. Pathak, pp 129-138 (World Scientific, 2000)
\bibitem{13} S. C. Tiwari, Fresh look on relativistic time and lifetime of an unstable particle, in Proceedings of Physical 
Interpretation of Relativity Theory, British Society for the Philosophy of Science (1988)
\bibitem{14} S. Weinberg, The Quantum Theory of Fields Volume 1 (C. U. P. 1995)
\bibitem{15} S. C. Tiwari, Symmetry and geometry: pursuit of beauty in physics, Contemporary Physics, 53, 485 (2012)
\bibitem{16} E. Nelson, Phys. Rev. 150, 1079 (1966)
\bibitem{17} L. de la Pena, A. M. Cetto and A. Valdes-Hernandez, Quantum behavior derived as an essentially stochastic phenomenon, Phys. Scripta
T151, 014008 (2012)
\bibitem{18} S. C. Tiwari, On the Schroedinger equation, Phys. Essays, 2, 31 (1989)
\bibitem{19} S. C. Tiwari, Coulomb- quantum oscillator correspondence in two dimension, pure gauge field and
half-quantized vortex, arXiv: 1902.02622 v1 [physics.gen-ph] ; Published in Mod. Phys. Lett. A
\bibitem{20} A. M. Cetto, L. de la Pena and A. Valdes-Hernandez, Proposed physical explanation for the electron spin and related antisymmetry,
Quantum Studies: Math and Found. 2017
\end{thebibliography}
\end{document}